\documentclass[fleqn,usenatbib]{mnras}
\usepackage{newtxtext,newtxmath,times,verbatim}
\usepackage[T1]{fontenc}
\DeclareRobustCommand{\VAN}[3]{#2}
\let\VANthebibliography\thebibliography
\def\thebibliography{\DeclareRobustCommand{\VAN}[3]{##3}\VANthebibliography}
\usepackage{graphicx}
\usepackage{amsmath}
\usepackage{ulem}
\usepackage{comment}
\usepackage{color}

\title[Quenching in dwarf galaxies]{The quenching of star formation in dwarf galaxies: new perspectives from deep-wide surveys}

\author[S. Kaviraj et al.]{S. Kaviraj,$^{1}$\thanks{E-mail: s.kaviraj@herts.ac.uk} I. Lazar,$^{1}$ A. E. Watkins,$^{1}$ C. Laigle,$^{2}$ G. Martin,$^{3,4,5}$ and R. A. Jackson$^{6}$\\
$^{1}$Centre for Astrophysics Research, University of Hertfordshire, Hatfield, AL10 9AB, UK\\
$^{2}$Institut d’Astrophysique de Paris, UMR 7095, CNRS, and Sorbonne Université, 98 bis boulevard Arago, 75014 Paris, France\\
$^{3}$School of Physics and Astronomy, University of Nottingham, University Park, Nottingham NG7 2RD, UK\\
$^{4}$Korea Astronomy and Space Science Institute, 776 Daedeokdae-ro, Yuseong-gu, Daejeon 34055, Korea\\
$^{5}$Steward Observatory, University of Arizona, 933 N. Cherry Ave, Tucson, AZ 85719, USA\\
$^{6}$Department of Physics and Astronomy, University of Victoria, Victoria, BC, Canada V8P 5C2\\
}

\begin{document}
\label{firstpage}
\pagerange{\pageref{firstpage}--\pageref{lastpage}}
\maketitle

\begin{abstract}
Dwarf galaxies dominate the galaxy number density, making them critical to our understanding of galaxy evolution. However, typical dwarfs are too faint to be visible outside the very local Universe in past surveys like the SDSS, which offer large footprints but are shallow. Dwarfs in such surveys have relatively high star formation rates, which boost their luminosity, making them detectable in shallow surveys, but also biased and potentially unrepresentative of dwarfs as a whole. Here, we use deep data to perform an unbiased statistical study of $\sim$7,000 nearby ($z<0.25$) dwarfs (10$^8$ M$_{\odot}$ < $M_{\star}$ < 10$^{9.5}$ M$_{\odot}$) in the COSMOS field which, at these redshifts, is a relatively low-density field. At $z\sim0.05$, $\sim$40 per cent of dwarfs in low-density environments are red/quenched, falling to $\sim$30 per cent by $z\sim0.25$. 
Red dwarfs reside closer to nodes, filaments and massive galaxies. Proximity to a massive galaxy appears to be more important in determining whether a dwarf is red, rather than simply its distance from nodes and filaments or the mean density of its local environment. Interestingly, around half of the red dwarfs reside outside the virial radii of massive galaxies and around a third of those also inhabit regions in the lower 50 per cent in density percentile (i.e. regions of very low ambient density). Around half of the red dwarf population is, therefore, quenched by mechanisms unrelated to environment, which are likely to be internal processes such as stellar and AGN feedback. 
\end{abstract}




\section{Introduction}
Dwarf galaxies dominate the galaxy number density in all environments and at all epochs \citep[e.g.][]{Wright2017,Martin2019}, making them central to a complete understanding of galaxy evolution. Much of our current knowledge of this regime comes from dwarf studies in our local neighbourhood, such as in the Local Group \citep[e.g.][]{Tolstoy2009} and around nearby massive galaxies \citep[e.g.][]{Poulain2021,Mao2021}. However, as we show in Section \ref{sec:deep_wide}, typical dwarfs are not bright enough to be detectable, outside the very local Universe, in past wide-area surveys like the SDSS \citep{Alam2015}, which offer large footprints but are comparatively shallow. The subset of dwarfs that do appear in such datasets have relatively high star formation rates (SFRs), usually driven by interactions \citep[e.g. Figure 1 in][]{Jackson2021a}. While these high SFRs boost their luminosity, making them detectable in shallow surveys, it also makes these galaxies biased and potentially unrepresentative of the dwarf population as a whole.  

This bias could manifest itself in multiple ways. For example, the fraction of red or quenched dwarfs could be underestimated in shallow surveys, because redder galaxies, which are fainter at a given stellar mass, will preferentially move out of the selection at progressively higher redshifts. Recent work using the SDSS has suggested that, below a stellar mass of $10^9$ M$_{\odot}$, the quenched fraction in nearby ($z<0.055$) dwarfs in low-density environments which reside at significant distances (> 1.5 Mpc) from the nearest massive galaxy may be close to zero \citep[e.g.][]{Geha2012}, a result which shows some tension with simulations \citep[e.g.][]{Sharma2022,Feldmann2023,Herzog2023}. It is worth considering whether at least some of this result could be driven by the inability of the SDSS to detect quenched dwarfs outside the very local Universe because they are too faint.   

In this context, it is interesting to note that when an analysis of SDSS dwarfs is performed at much lower redshift, which reduces but does not completely eliminate the bias, the red/quenched fraction appears to be higher. For instance, \citet{Barazza2006} have shown that SDSS dwarfs with stellar masses less than $\sim$10$^{8.5}$ M$_{\odot}$ at $z<0.02$ exhibit a pronounced red sequence. Since red galaxies are typically dominated by systems that are quenched \citep[e.g.][]{Kaviraj2007}, the quenched fraction in dwarfs suggested by this study is significantly higher than zero. Note, however, that these dwarfs are not selected to be at large distances from massive galaxies, which could partly explain the discrepancy. Nevertheless, the conclusions of Barazza et al. appear consistent with \citet{Tanoglidis2021}, \citet{Thuruthipilly2024} and \citet{Lazar2024a,Lazar2024b}, who have studied nearby low-surface-brightness and dwarf galaxies using surveys that are significantly deeper than the SDSS and also find significant populations of red dwarfs in low-density environments out to moderate redshift.  

The identification of interesting sub-populations may also be affected by the biased nature of the dwarfs in shallow surveys. For example, identifying dwarfs which host AGN, using common techniques that attempt to separate AGN from star-forming systems -- e.g. the emission-line `BPT' method  \citep{Baldwin1981,Veilleux1987} -- may be complicated by the fact that dwarfs detected by shallow surveys are typically star-forming. The contribution of star formation to diagnostic features like emission lines may swamp the contribution due to the AGN. This is particularly true when single fibre spectroscopy which covers a significant fraction of the galaxy is utilised, resulting in spuriously low AGN fractions \citep[e.g.][see also \citet{Davis2022,Bichanga2024}]{Mezcua2024}. In summary, results of past dwarf-galaxy studies that are underpinned by data from shallow surveys need to be interpreted in the context of these biases. Furthermore, accurate statistical estimates of key properties like red and quenched fractions in the dwarf regime benefit greatly from surveys that are both deep and wide. 

Overall, the issues described above have two important consequences for our understanding of galaxy evolution. First, our observational picture of how galaxies form and evolve over cosmic time is overwhelmingly dominated by bright (i.e. massive) galaxies. Second, our theoretical models are largely calibrated to reproduce the properties of massive galaxies (only). Our current understanding of the physics of galaxy evolution is, therefore, likely to be incomplete. In this context, it is worth noting that, given their shallower potential wells, the impact of many key processes, such as baryonic feedback and tidal perturbations, is likely to be stronger in dwarfs than in massive galaxies. This makes dwarfs more sensitive laboratories for studying such processes \citep[e.g.][]{Martin2019,Martin2021,Jackson2021b,Watkins2023,Uzeirbegovic2024}, in order to gain a better understanding of how they shape galaxy evolution. 

Some recent surveys, such as MATLAS \citep{Duc2015}, SAGA \citep{Geha2017}, LIGHTS \citep{Trujillo2021} and the Fornax Deep Survey \citep{Venhola2018}, reach the depths required to assemble mass-complete samples of relatively luminous nearby dwarfs. In these surveys, dwarfs are identified based on their proximity to massive systems. However, as we show below, proximity to massive galaxies is potentially the most significant factor in making dwarf galaxies red. In addition, redder dwarfs tend to exhibit higher fractions of systems with early-type morphology \citep[e.g.][]{Lazar2024a}. Thus, while these surveys are able to probe more complete samples of dwarfs, the dwarf population may be, by construction, skewed towards relatively high-density environments in which dwarfs show biases in terms of colour and morphology.  

A systematic study of dwarfs in low-density environments requires surveys that are deep, wide and not centred either on massive galaxies or known regions of high density. The advent of several such surveys -- such as the Hyper Suprime-Cam Subaru Strategic Program \citep[HSC-SSP,][]{Aihara2018a}, the forthcoming Legacy Survey of Space and Time (LSST) via the Rubin Observatory \citep[e.g.][]{Ivezic2019,Watkins2024} and Euclid \citep{Laureijs2011} -- is poised to revolutionise our understanding of galaxy evolution \citep[e.g.][]{Kaviraj2020}, by offering large unbiased samples of dwarfs outside the very local Universe. Together with deep-wide surveys in other wavelengths, such as those using the Roman telescope in the infrared \citep[e.g.][]{Spergel2015} and the SKA in the radio \citep[e.g.][]{Braun2015}, this new generation of datasets will enable us to explore, for the first time, aspects of galaxy evolution in the dwarf regime that we were previously restricted to studying only in massive galaxies.

Here, we use deep-wide data in the COSMOS field to perform a statistical study of star formation and quenching in $\sim$7,000 dwarf (10$^8$ M$_{\odot}$ < $M_{\star}$ < 10$^{9.5}$ M$_{\odot}$) galaxies in the nearby ($z<0.25$) Universe. Our study provides both an unbiased exploration of dwarfs outside the local neighbourhood and a demonstration of the novel analyses that can be performed using the new and forthcoming surveys that will dominate the astronomical landscape in the next decades. 

This paper is organised as follows. In Section \ref{sec:data}, we outline the datasets that underpin this study. In Section \ref{sec:deep_wide}, we explore why deep-wide surveys are necessary for such an analysis in the first place. We study the detectability and completeness of dwarf populations as a function of stellar mass and redshift and quantify the differences between dwarfs that appear in deep-wide surveys and those that are detectable in shallow datasets like the SDSS. In Sections \ref{sec:colours} and \ref{sec:sfms}, we calculate the fraction of dwarfs that reside on the optical red sequence, consider how the star formation main sequence extends into the dwarf regime and estimate the fraction of dwarfs that are quenched. In Section \ref{sec:quenching}, we study the distances of red and blue dwarfs from nodes, filaments and massive galaxies in order to explore the processes that drive the quenching of star formation in nearby dwarf galaxies. We also compare our results to the findings of recent theoretical work. We summarise our findings in Section \ref{sec:summary}. 


\section{Data}
\label{sec:data}

\subsection{Physical parameters and visual inspection to remove spurious sources}

We use physical parameters (photometric redshifts, stellar masses, rest-frame colours and SFRs) from the Classic version of the COSMOS2020 catalogue \citep{Weaver2022}, a high-precision value-added catalogue of sources in the 2 deg$^2$ region of the COSMOS field and a successor to the benchmark COSMOS2015 dataset \citep{Laigle2016} which was created using similar techniques. The physical properties in COSMOS2020 are calculated using deep, multi-wavelength UV to mid-infrared photometry in 40 broad and medium band filters from several instruments: GALEX \citep{Zamojski2007}, MegaCam/CFHT \citep{Sawicki2019}, ACS/HST \citep{Leauthaud2007}, Subaru/Hyper Suprime-Cam \citep{Aihara2019}, Subaru/Suprime-Cam \citep{Taniguchi2007,Taniguchi2015}, VIRCAM/VISTA \citep{McCracken2012} and IRAC/Spitzer \citep{Ashby2013,Steinhardt2014,Ashby2015,Ashby2018}. 

A particular novelty is the incorporation of optical ($grizy$) data from the Ultra-deep layer of the HSC-SSP, which has a point source depth of $\sim$28 mag \citep{Aihara2019}, for object detection. The detection limit of HSC-SSP Ultra-deep is $\sim$5 mag fainter than standard depth SDSS imaging and $\sim$10 mag fainter than the magnitude limit of the SDSS spectroscopic main galaxy sample (MGS). As we explore in Section \ref{sec:deep_wide}, this facilitates the identification of unbiased samples of dwarfs that are likely to be complete down to a stellar mass of $M_{\star}$ $\sim$ 10$^{8}$ M$_{\odot}$ and out to at least $z\sim0.3$. It is worth noting that SDSS photometry is typically not deep enough to extract reliable photometric redshifts, particularly for faint objects like dwarf galaxies. Most scientific analyses that use the SDSS dataset, including past dwarf-galaxy studies, employ the MGS, which is restricted to galaxies with $r<17.77$ \citep{Strauss2002},  $\sim$5 mag brighter than the nominal depth of standard SDSS imaging. 

The survey images are homogenized to a common point-spread function, and fluxes are extracted within circular apertures, using the \textsc{SExtractor} and \textsc{IRACLEAN} codes for UV/optical and infrared photometry respectively. Physical parameters are calculated using the \textsc{LePhare} SED-fitting algorithm \citep{Arnouts2002,Ilbert2006}\footnote{We direct readers to Section 5.1 in \citet{Weaver2022} for a description of the SED fitting. Briefly, the template library spans all galaxy morphological types \citep{Polletta2007}, blue star-forming models from \citet{Bruzual2003} and templates that account for AGN \citep[e.g.][]{Salvato2009,Salvato2011}. Extinction is included as a free parameter, with a flat prior in the reddening ($E_{\rm B-V}$ < 0.5). Emission lines are added using the relationship between the UV luminosity and [O II] emission-line flux, following \citet{Ilbert2009}.}. Photometric redshifts have typical accuracies better than 1 and 4 per cent for bright ($i<22.5$ mag) and faint ($25<i<27$ mag) galaxies respectively, making this catalogue well-suited for our purposes. 

To construct our sample, we first select objects that are classified as galaxies by \textsc{LePHARE} (`type' = 0 in the COSMOS2020 catalogue) and are also classified as `extended' (i.e. galaxies) in the HSC $griz$ filters. We then restrict our study to galaxies which have stellar masses in the range 10$^{8}$ M$_{\odot}$ < $M_{\rm{\star}}$ < 10$^{9.5}$ M$_{\odot}$, redshifts in the range $z<0.25$, lie within the HSC-SSP footprint and outside masked regions and have both $u$-band and mid-infrared photometry (since a wide wavelength baseline aids the accuracy of the parameter estimation e.g. \citet{Ilbert2006}). This produces an initial sample of $\sim$7,000 low-mass galaxies. 

\begin{figure}
\center
\includegraphics[width=\columnwidth]{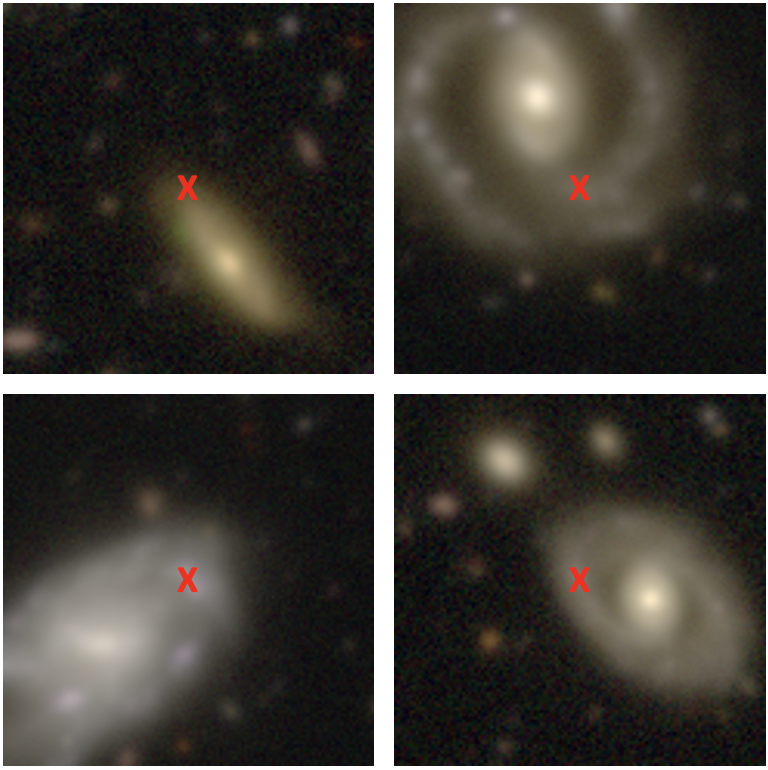}
\caption{Examples of four objects which are classified as low-mass galaxies but which are actually regions of massive galaxies that have been shredded by the deblender. The position of the HSC object is shown using a red cross. Around 0.8 per cent of objects classified as dwarfs in the COSMOS2020 catalogue fit this description and have been removed from our analysis.}
\label{fig:visual_inspection}
\end{figure}

As noted in \citet{Davis2022}, a small fraction of objects identified as low-mass galaxies in such catalogues are actually regions within nearby massive galaxies that have been shredded by the deblender. Therefore, we visually inspect the HSC image of each low-mass object in the initial sample described above, in order to identify and remove such spurious sources (see Figure \ref{fig:visual_inspection} for examples). In line with the findings of \citet{Davis2022}, these account for $\sim$0.8 per cent of objects. We remove these objects from our final sample. 

It is worth noting here that large, unbiased spectroscopic samples of dwarfs outside the very local Universe (particularly in low density environments), will be difficult to assemble in the near term, given the limits of current and forthcoming instrumentation. Photometric studies, such as this one, are therefore likely to form the basis of statistical studies of dwarf galaxies, outside the local neighbourhood, for the forseeable future. 

\subsection{DisPerSE: measurement of local density and the location of galaxies in the cosmic web}
\label{sec:disperse}

We use DisPerSE \citep{Sousbie2011}, a structure-finding algorithm, to measure the local density and the locations of galaxies within the cosmic web, such as their distances to the nearest nodes and filaments. DisPerSE uses Delaunay tessellations to measure the density field, calculated using the positions of galaxies \citep{Schaap2000}. It then uses segments to connect topological saddle points with the local maxima in the density map, forming a set of ridges that constitute a `skeleton', which describes the network of filaments that define the cosmic web. Stationary points in the density map -- i.e. minima, maxima and saddles -- correspond to the locations of voids, nodes and the centres of filaments respectively. We refer readers to \citet{Sousbie2011} for further details of the algorithm. 

\begin{figure}
\center
\includegraphics[width=\columnwidth]{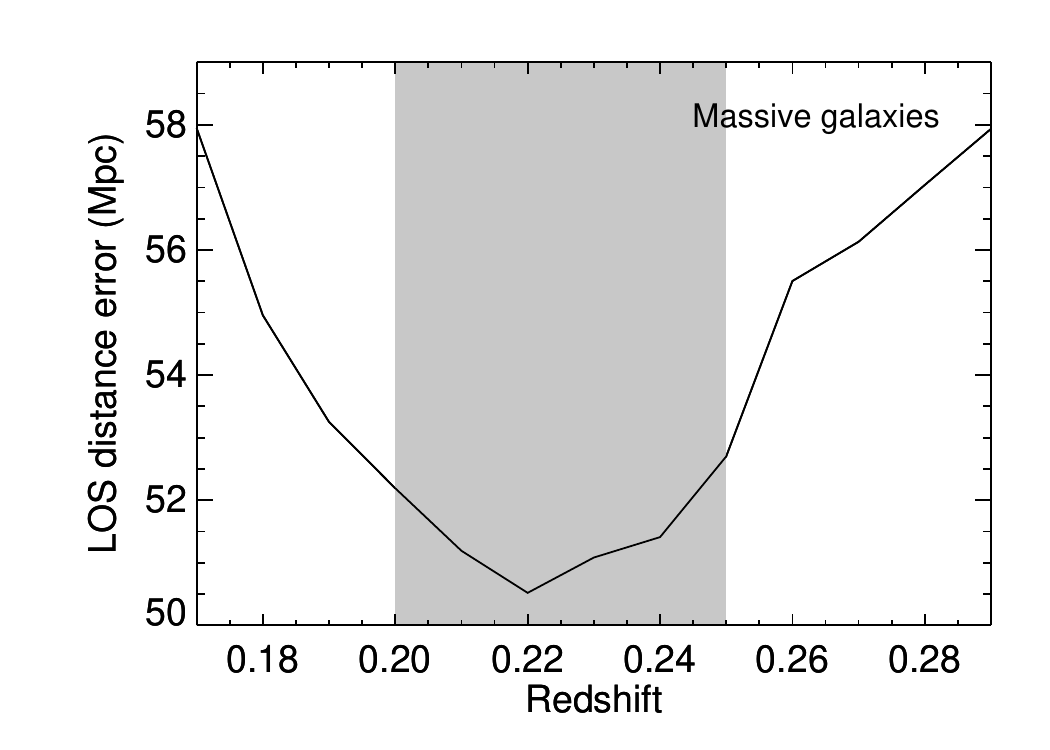}
\caption{The median 1$\sigma$ redshift uncertainty of massive ($M_{\star}$ > 10$^{10}$ M$_\odot$) galaxies converted to a line-of-sight (LOS) comoving distance error in Mpc, as a function of redshift. Note that, although we show this figure out to $z=0.29$, the redshift limit of our study is $z=0.25$. Only galaxies with $M_{\star}$ > 10$^{10}$ M$_\odot$ are used to construct density maps, as they have the most accurate redshifts and will dominate the local gravitational potential well. We restrict the density analysis, in Section \ref{sec:quenching}, to the grey shaded region, which exhibits a broad minimum in the median redshift error and the corresponding LOS distance uncertainty. The uncertainties in the LOS distances in this region are around 3 per cent.} 
\label{fig:LOS_uncertainty}
\end{figure}

The properties of the skeleton are determined by a `persistence' parameter, which sets a threshold value for defining pairs of critical points within the density map. A persistence of \textit{N} results in a skeleton where all critical pairs with Poisson probabilities below \textit{N}$\sigma$ from the mean are removed. We follow the methodology of \citet{Laigle2018}, who have implemented DisPerSE on redshift slices of similar widths as in our analysis, constructed from the COSMOS2015 \citep{Laigle2016} catalogue. The same methodology has also recently been used to perform a similar density analysis using the COSMOS2020 catalogue by \citet{Lazar2023} and \citet{Bichanga2024}. 

The thickness of the slices is driven by the redshift uncertainties of the galaxy sample. \citet{Laigle2018} have shown, using the Horizon-AGN cosmological simulation \citep{Dubois2014,Kaviraj2017}, that datasets like COSMOS2015 which offer high photometric redshift precision (and, by extension, COSMOS2020 which provides similar precision) can recover the broad 3D properties of the cosmic web from 2D projected density maps. Following \citet{Laigle2018} and \citet{Lazar2023}, we use a persistence of 2 in this study, which removes ridges close to the noise level, where structures could be spurious. To correctly estimate the topology of galaxies close to the field boundary, a surface of `guard' particles is added outside the boundary coordinates, with new particles added by interpolating the calculated density at the boundary itself. Additionally, we exclude galaxies which lie within 0.1 degrees of the boundary from our analysis in order to completely avoid any edge effects. 

The accuracy of the COSMOS2020 redshifts enables us to employ well-defined and relatively narrow redshift slices to build our density maps. We only use massive ($M_{\star}$ > 10$^{10}$ M$_{\odot}$) galaxies to build these maps, as they have the smallest redshift errors and will dominate the local gravitational potential wells. When constructing each density map, individual galaxies are weighted by the area under their redshift probability density function that is contained within the slice in question. This takes into account the fact that the photometric redshifts of massive galaxies, although very accurate in COSMOS2020, do have associated errors. Figure \ref{fig:LOS_uncertainty} indicates the {\color{black}line-of-sight} comoving distance errors that correspond to the median 1$\sigma$ redshift uncertainties of massive galaxies, as a function of redshift. For our sample, the 1$\sigma$ redshift uncertainty reaches a broad minimum in the redshift range $0.2<z<0.25$, corresponding to physical distance errors between $\sim$50 -- 53 Mpc. In addition, the number of massive galaxies at $z<0.2$ is not large enough to construct density maps. Thus, in Section \ref{sec:quenching}, we restrict the analysis of environmental properties to the redshift range $0.2<z<0.25$.  

Since we are interested in the role of environment in influencing the evolution of dwarfs, it is instructive to consider the types of large-scale structures that are likely to exist within our COSMOS2020 footprint. We note first that the COSMOS field is not centred on a known region of high density. In the redshift range that we probe for the environmental analysis, the COSMOS2020 footprint corresponds to a linear transverse size of $\sim$26 Mpc. To explore the types of structures that might be present in a field of this size, we consider the NewHorizon cosmological simulation \citep{Dubois2021}, which has a similar (albeit slightly smaller) size to the COSMOS2020 field at these redshifts. The largest dark matter halo in NewHorizon has a mass of $\sim$10$^{13}$ M$_{\odot}$, similar to that of a relatively large group\footnote{In comparison, a small cluster like Fornax has a virial mass of $\sim$7$\times$10$^{13}$ M$_{\odot}$ \citep{Drinkwater2001}, while larger clusters like Virgo and Coma have virial masses of $\sim$10$^{15}$ M$_{\odot}$ \citep[e.g.][]{Fouque2001,Gavazzi2009}.}. Thus, the large scale structure within the COSMOS2020 field at these redshifts is likely to contain relatively low-density environments, comprising groups of galaxies and systems in the field. 


\section{The need for deep-wide surveys: detectability and completeness in the dwarf regime}
\label{sec:deep_wide}

We first consider why deep-wide surveys are important for performing unbiased statistical studies of dwarfs outside the very local Universe. We begin, in Figure \ref{fig:different_surveys}, by illustrating this point visually. We show optical images of several dwarfs, which have stellar masses and redshifts close to the median values of our sample ($\sim$10$^{8.5}$ M$_{\odot}$ and $z\sim0.19$) in three surveys of varying depths\footnote{The images are created using the cutout tool created by Yao-Yuan Mao, which can be found at this address: https://yymao.github.io/decals-image-list-tool/.}. The top three rows show examples of red dwarfs (rest-frame $(g-i)$ > 0.7), while the bottom three rows show examples of blue dwarfs (rest-frame $(g-i)$ < 0.7). While the dwarfs are well-detected in the HSC-SSP Ultra-deep images (left panel), which are $\sim$4 mag deeper than standard-depth SDSS imaging, they are less reliably detected in the DECaLS survey \citep[middle panel;][]{Dey2019}, which is $\sim$2 mag deeper than the SDSS and close to the noise, especially for the red galaxies, in the SDSS itself (right panel). It is worth noting here that none of these objects are bright enough to have an SDSS spectrum and therefore they do not appear in the MGS at all. 

{\color{black}We study this point more precisely by quantifying the completeness of dwarf galaxy populations in these different surveys, as a function of stellar mass and redshift. At a given redshift, a population of galaxies with a given stellar mass can be considered complete, and therefore unbiased, if \textit{any} galaxy with that stellar mass is detectable in the survey in question. We can estimate the redshift out to which galaxies of a given stellar mass are complete by considering an object which is likely to have the faintest magnitude at that stellar mass. If this faintest `limiting case' is brighter than the detection limit of the survey in question, then all galaxies with that stellar mass will, in principle, also be detectable. 

Here, we define this faintest limiting case to be a purely-old `simple stellar population' (SSP), that forms in an instantaneous burst at $z=2$. Since virtually no real galaxy at low redshift is composed of a purely-old stellar population, this (hypothetical) object allows us to explore the completeness of different surveys. We consider SSPs with two metallicities, 0.05 Z$_{\odot}$ and 0.5 Z$_{\odot}$, constructed using the \citet{Bruzual2003} stellar models. The broad conclusions are not strongly sensitive to the chosen metallicity, as we show below. The local stellar mass -- metallicity relation \citep[e.g.][]{Gallazzi2005} indicates that the stellar metallicity of dwarfs at the median stellar mass of our sample is $\sim$0.2 Z$_{\odot}$. The chosen metallicities therefore bracket the metallicities of nearby dwarf galaxies in our stellar mass range of interest.  

In the top panel of Figure \ref{fig:detectability}, we use these SSPs to estimate the redshifts at which galaxy populations of different stellar masses are complete in different surveys. {\color{black}We perform this exercise in two ways, first considering just the total magnitude and second considering the effective surface brightness (the conclusions are similar, regardless of the method used). For the latter, we combine the total magnitude with the 95th percentile value of the effective radius distribution of the dwarfs in our study, in order to calculate a limiting value for the effective surface brightness. We then estimate the redshift at which either the magnitude or surface brightness of our faintest limiting case equals the magnitude or surface brightness limit of the surveys in question, as a function of stellar mass.}

The curves show these redshifts for the following surveys: the HSC-SSP Ultra-Deep (black) and Wide (blue) layers, the new DESI Bright Galaxy Sample \citep[orange;][]{Hahn2023} and the SDSS MGS (red). For each survey, the solid and dashed lines indicate the detection thresholds calculated using our purely-old SSPs with metallicities of 0.05 Z$_{\odot}$ and 0.5 Z$_{\odot}$ respectively. Curves calculated using magnitudes and surface brightnesses are indicated using `mag' and `SB' respectively. {\color{black} The conclusions derived from total magnitudes and effective surface brightnesses are similar because, at our redshifts of interest, the sizes of dwarfs are small (see Appendix \ref{sec:effective_radii}). We note that these completeness curves are likely to be pessimistic because, as Figure \ref{fig:red_fraction_summary} indicates, very few nearby dwarfs are actually consistent with purely old stellar populations. In other words, it is likely that our dwarf populations are complete out to higher redshifts than those indicated by these curves}. 

In the context of these completeness thresholds, galaxy populations are complete (and unbiased) in the part of the redshift -- stellar mass parameter space which is below the curve in question. Note that this does not mean that no galaxies above the curves will exist in a particular survey. Rather, galaxy populations that are above and further away from these curves will be progressively more biased towards bluer galaxies (and against redder objects). For example, at $z\sim0.15$, our purely-old limiting case, even at 10$^7$ M$_{\odot}$, is brighter than the detection limit of the HSC-SSP Ultra-deep survey, indicating that the dwarf population should be complete down to this mass limit at this redshift. Similarly, at $z\sim0.3$ the population is likely to be complete down to 10$^{7.5}$ M$_{\odot}$. The galaxy population that underpins our study is indicated using the grey rectangle. {\color{black}Comparison of this grey region to the black lines indicates that this population is complete within our mass and redshift ranges of interest, indicating, therefore, that our results will not be biased by subsets of dwarfs (e.g. redder galaxies) moving out of the selection.}

\begin{figure}
\center
\includegraphics[width=\columnwidth]{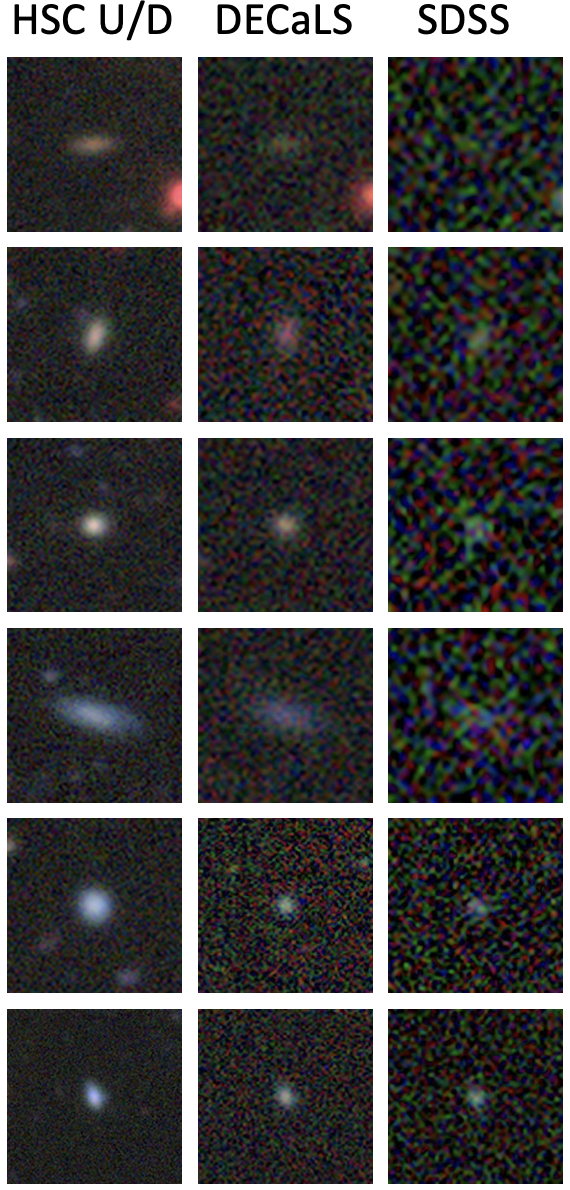}
\caption{Images from surveys of varying depths, of several dwarfs which have stellar masses and redshifts close to the median values in our sample ($\sim$10$^{8.5}$ M$_{\odot}$ and z$\sim$0.19 respectively). The top three rows show examples of red dwarfs (rest-frame $(g-i)$ > 0.7), while the bottom three rows show examples of blue dwarfs (rest-frame $(g-i)$ < 0.7). While the dwarfs are well-detected in the HSC-SSP Ultra-deep images (left), which are around four mag deeper than standard-depth SDSS imaging, they are less reliably detected in the DECaLS survey (middle), which is two mag deeper than SDSS and often close to the noise (especially for the red galaxies) in the SDSS itself (right). Note that none of these objects are bright enough to have an SDSS spectrum and therefore they do not appear in the MGS.}
\label{fig:different_surveys}
\end{figure}

\begin{figure}
\center
\includegraphics[width=\columnwidth]{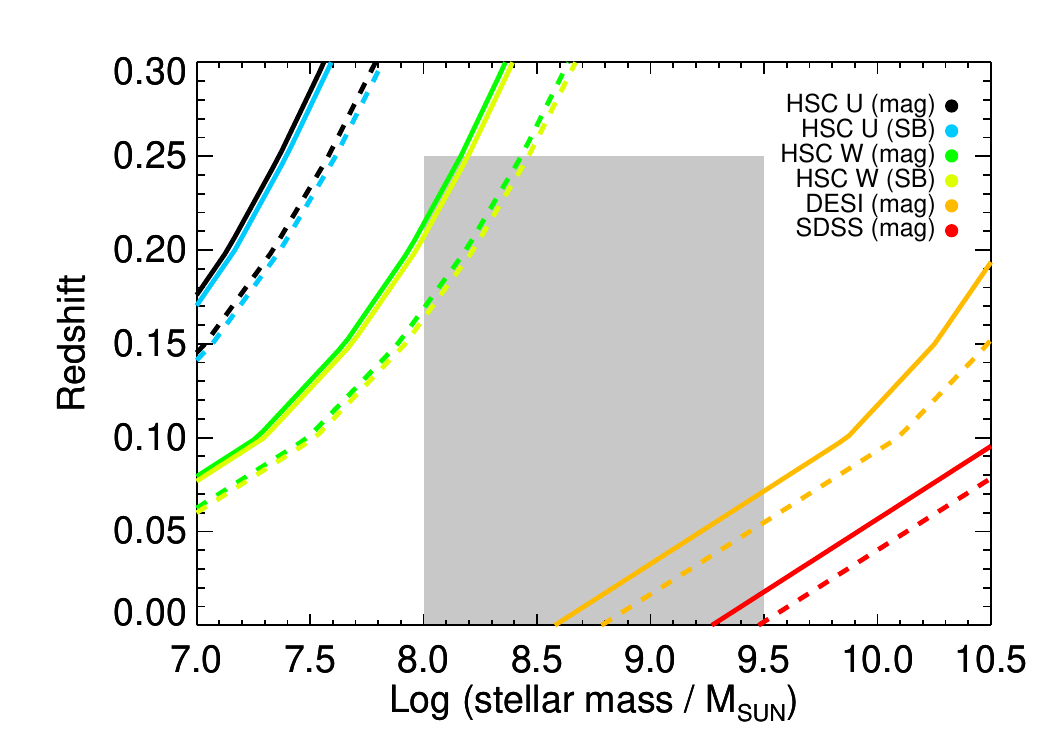}\\
\includegraphics[width=\columnwidth]{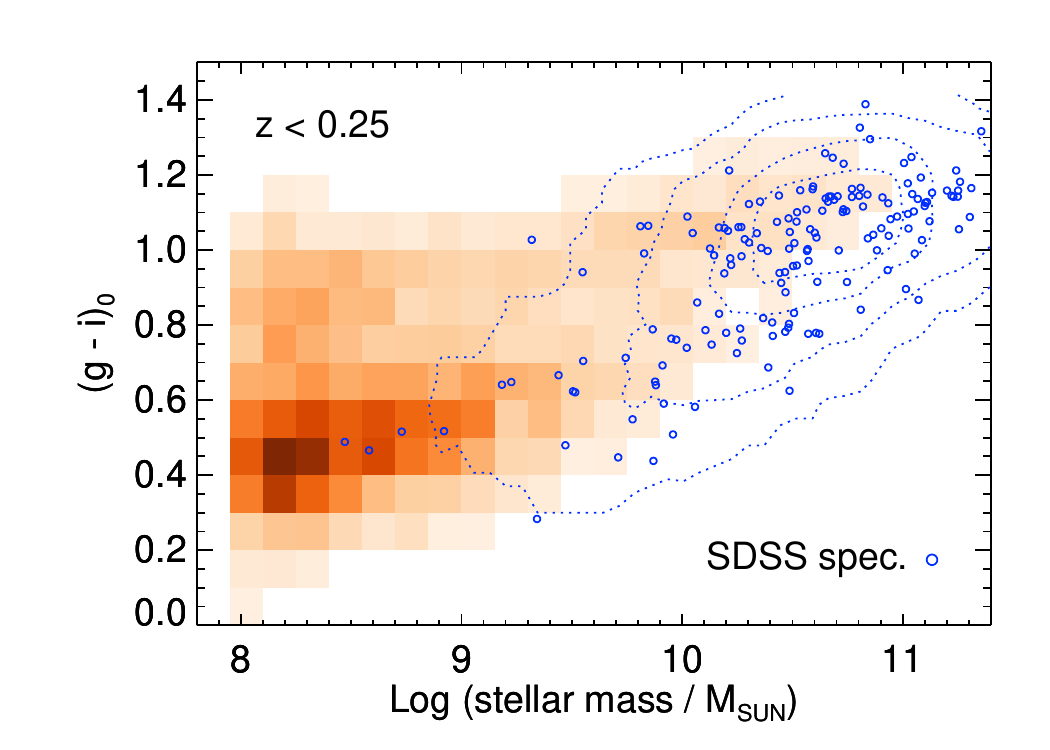}
\caption{\textbf{Top:} Redshifts at which galaxy populations of different stellar masses are complete in various surveys: the HSC-SSP Ultra-Deep (black and blue) and Wide (green and yellow) layers, the new DESI Bright Galaxy Sample \citep[orange][]{Hahn2023} and the SDSS MGS (red). For each survey, the solid and dotted lines indicate detection thresholds calculated using purely-old SSPs with metallicities of 0.05 Z$_{\odot}$ and 0.5 Z$_{\odot}$ respectively. {\color{black}Curves calculated using magnitudes and surface brightnesses are indicated using `mag' and `SB' respectively.} Galaxy populations are complete in the part of the parameter space which is below the curve in question. The galaxy population that underpins our study is indicated using the grey rectangle. \textbf{Bottom:} The rest-frame $(g-i)$ colour of the COSMOS2020 galaxies, shown as a heatmap, with the location of SDSS spectroscopic objects within the COSMOS2020 footprint shown using blue contours. A random sample of these galaxies is overplotted using the blue open circles. While massive galaxies appear in the SDSS spectroscopic sample regardless of whether they are red or blue, galaxies become progressively bluer with decreasing stellar mass, as red objects preferentially fall out of the selection because they are fainter at a given stellar mass.}
\label{fig:detectability}
\end{figure}

\begin{figure*}
\center
\includegraphics[width=\columnwidth]{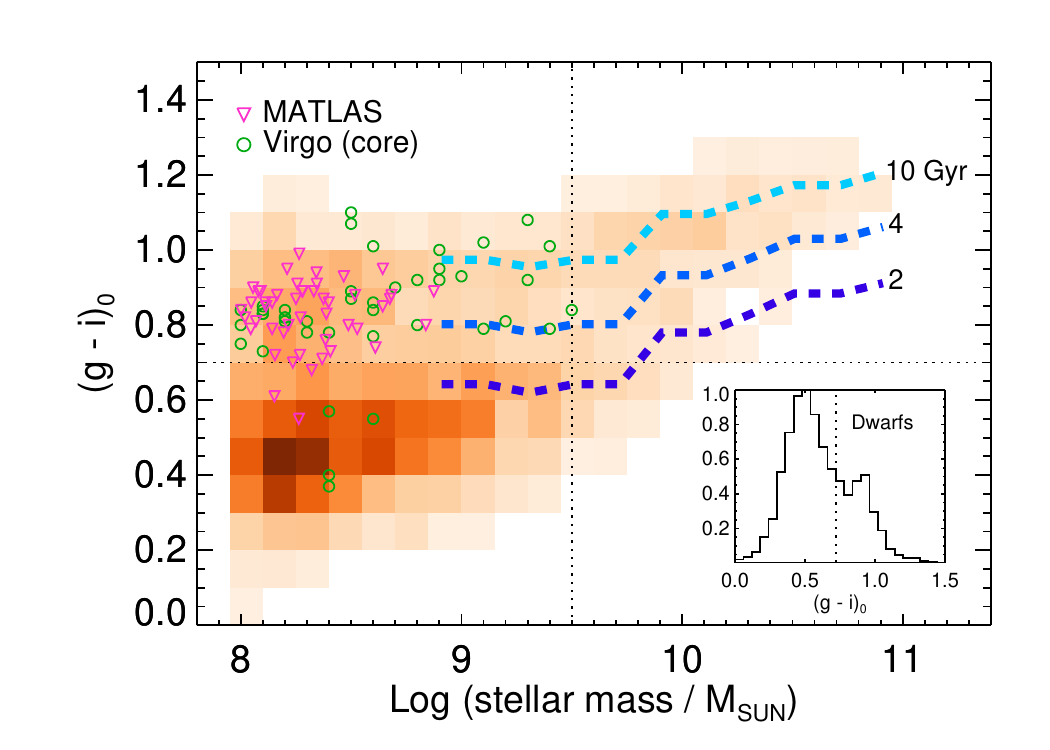}
\includegraphics[width=\columnwidth]{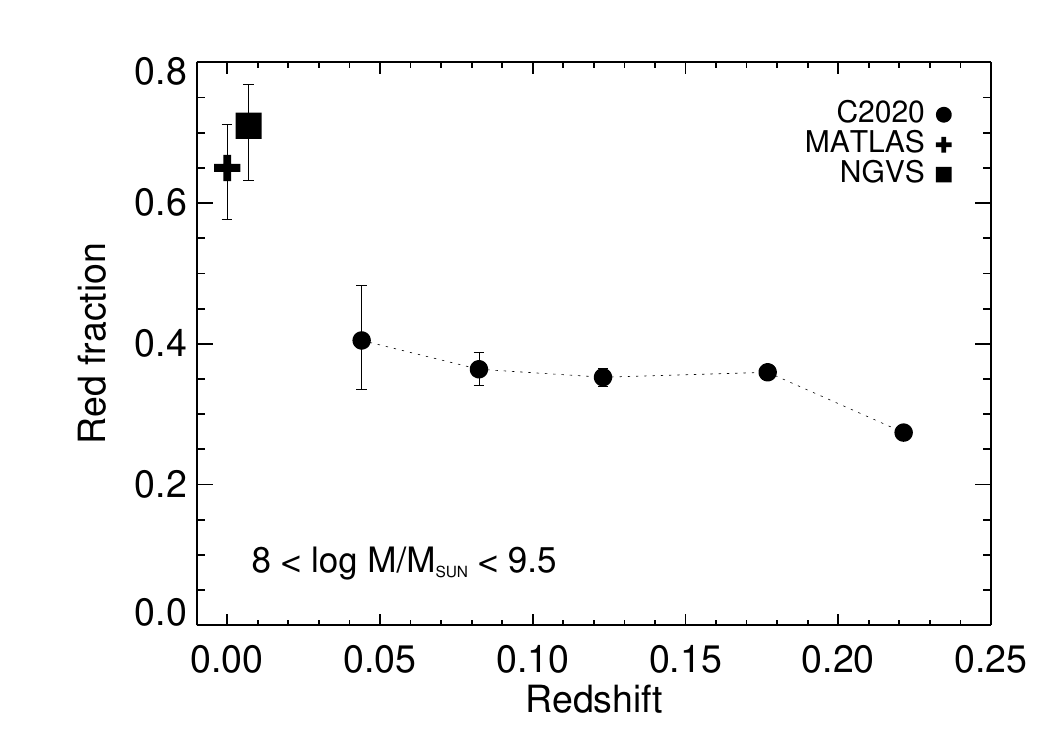}
\caption{\textbf{Left:} Rest-frame $(g-i)$ colours of the COSMOS2020 galaxies, shown as a heatmap, with those in the MATLAS survey and in the core of Virgo shown using pink and green symbols respectively. The thick dashed lines show SSPs which form at various look-back times (10, 4, and 2 Gyrs). Each SSP has the median metallicity, at a given stellar mass, from the mass-metallicity relation in \citet{Gallazzi2005}. {\color{black} The variation in the rest-frame colour is driven by the variation in the median metallicity with stellar mass. The dotted horizontal line shows the demarcation between red and blue galaxies at rest-frame $(g-i)=0.7$, following \citet{Lazar2024a}, which is based on the fact that the 2D distribution of dwarf galaxies in the colour -- stellar mass plane is bimodal around this value. The dotted vertical line indicates the upper stellar mass limit for our definition of dwarf galaxies. The inset shows the distribution of rest-frame $(g-i)$ colours for the dwarf population only.} \textbf{Right:} The evolution of the red fraction (i.e. the fraction of galaxies with $(g-i)>0.7$) for the COSMOS2020 dwarfs. Shown overplotted are red fractions in dwarfs in the MATLAS survey and in the core of Virgo from the NGVS survey.} 
\label{fig:red_fraction_summary}
\end{figure*}

The situation for the SDSS spectroscopic MGS (red solid and dotted lines), which is the basis for many past dwarf studies, is very different. Only at the uppermost end of the dwarf mass range, and at very low redshift ($z<0.02$), are galaxy populations likely to be complete in the SDSS. In other words, typical galaxies with M$_*$ < 10$^{9.5}$ M$_{\odot}$ are unlikely to appear in the SDSS spectroscopic sample outside the very local Universe. As noted above, dwarfs outside our local neighbourhood will generally appear in shallow surveys like the SDSS only if they have high SFRs, which boosts their luminosity and makes them detectable in such shallow datasets. However, as we show in the bottom panel of this figure, these high SFRs also make these dwarfs blue and potentially unrepresentative of the dwarf population as a whole. 

The bottom panel of this figure shows the consequence of the incompleteness described above. The rest-frame $(g-i)$ colour vs stellar mass of galaxies in COSMOS2020 at $z<0.25$ is shown using the heatmap. The blue contours indicate the locations of SDSS spectroscopic objects, from the MPA-JHU catalogue \citep{Kauffmann2003,Brinchmann2004}, that exist within this footprint and in this redshift range. A random sample of SDSS spectroscopic objects are shown overplotted using open blue circles. The general trend indicates that, while massive galaxies appear in the SDSS MGS regardless of colour, galaxies become increasingly blue with decreasing stellar mass. At the lower end of the dwarf galaxy mass range (e.g. $M_{\star}$ < 10$^{9}$ M$_{\odot}$) the SDSS-detected systems only exhibit blue rest-frame colours. As we show in Section \ref{sec:colours} below, this bias likely translates into a spuriously low fraction of red (or quenched) dwarfs in such shallow surveys. 


\section{Red fractions in dwarf galaxies}
\label{sec:colours}

We begin our analysis of the properties of our dwarfs by considering their rest-frame colours, which are a powerful diagnostic of the recent star formation history of galaxies \citep[e.g.][]{Strateva2001,Pandey2024}. In the left-hand panel of Figure \ref{fig:red_fraction_summary}, we show the rest-frame $(g-i)$ colour of the COSMOS2020 population as a heatmap. We study this colour because it facilitates comparison with populations of dwarfs in local high-density regions, from Virgo and around nearby massive galaxies, which also have photometry in these filters. In the dwarf regime in the nearby Universe, the rest-frame $(g-i)$ colour is bimodal in the colour -- stellar mass plane around $(g-i)\sim 0.7$, with well-defined red and blue peaks \citep{Lazar2024a}. Overlaid are rest-frame $(g-i)$ colours of simple stellar populations (SSPs) which form at look-back times of 2, 4 and 10 Gyrs (taken from the \citet{Bruzual2003} population synthesis models). Since galaxies show a relationship between stellar mass and metallicity, we use SSPs which have the median metallicity at a given stellar mass, taken from the mass -- metallicity relation in \citet{Gallazzi2005}. While real galaxies do not have star formation histories that correspond exactly to SSPs, these SSP-based colours serve as a useful guide to understanding the broad features of the star-formation histories of galaxies in this colour-mass diagram. 

Dwarfs in the blue peak lie blueward of the 2 Gyr SSP, suggesting that these galaxies are likely to have had some star formation within the last 2 Gyrs. In the same vein, the red dwarfs are unlikely to have had significant star formation in the last 2 Gyrs. Indeed, most dwarfs in the red peak lie blueward of the 4 and 10 Gyr SSPs, indicating that very few dwarfs are consistent with having purely old stellar populations and a `monolithic' formation scenario. While the COSMOS footprint is likely to be dominated by low density environments at the redshifts probed by this study, it is instructive to compare the rest-frame colours of our dwarfs to those in relatively high-density environments. The green and pink symbols show dwarfs in the MATLAS \citep{Duc2015,Poulain2021} and NGVS \citep{Ferrarese2012} surveys. The former samples  environments around massive galaxies like the Milky Way, while the latter is a survey of galaxies in the Virgo cluster. Here, we specifically consider dwarfs in the core of Virgo from \citet{Sanchez-Janssen2019}, which are likely to be characteristic of the highest-density regions in the nearby Universe. 

In the right-hand panel of this figure we summarise the red fractions in the dwarf regime (i.e. the fraction of galaxies that lie redward of rest-frame $(g-i)=0.7$). At low redshift ($z\sim0.05$), $\sim$40 per cent of our dwarfs are red, with the red fraction falling slowly with redshift to below 30 per cent by $z\sim0.25$. Recall, from the discussion in Section \ref{sec:deep_wide}, that the dwarf population in this study is complete at these redshifts. Thus, this evolution is not due to red dwarfs falling out of the selection at progressively higher redshift but rather due to the real evolution in the star-formation properties of dwarfs across this period of cosmic time. The red fractions calculated using the SDSS MGS are underestimated by around a factor of 3 at $z<0.05$ and a factor of 8 at $z\sim0.08$. 

Our red fractions are consistent with the value ($\sim$35 per cent) derived by \citet{Lazar2024a} at $z<0.08$ using HSC data. It is also consistent with other recent work which has studied nearby low-surface-brightness galaxies (which are dominated by dwarfs) using relatively deep data from the Dark Energy Survey \citep[DES; e.g.][]{Tanoglidis2021,Thuruthipilly2024}. The red fractions from these studies, using the same $(g-i)$ colour as we use here, are around 30 per cent. Note, however, that the HSC data in this study is $\sim$3 magnitudes deeper than DES, which likely explains the small discrepancy in these red fractions. In summary, our results agree well with the findings of other work which uses relatively deep datasets. 

The red fraction is elevated around nearby massive galaxies in the MATLAS survey and in the core of Virgo by factors of $\sim$1.6 and $\sim$1.8 respectively, compared to the value in low-density environments. This is expected, since additional processes in relatively dense environments, such as tidal interactions \citep{Moore1998,Martin2019,Jackson2021a} and ram pressure stripping \citep[e.g.][]{Balogh2000,Hester2006}, will increase the removal of gas in dwarfs, reducing star formation activity even further and increasing the probability of dwarfs becoming red. However, the significant red fractions in dwarfs in low-density environments suggest that, in an appreciable fraction of dwarfs, the quenching of star formation does not rely on environmental factors but is likely to be triggered by internal processes like baryonic (stellar or AGN) feedback. We return to this point and explore it in more detail in Section \ref{sec:quenching} below. 

\begin{figure}
\center
\includegraphics[width=\columnwidth]{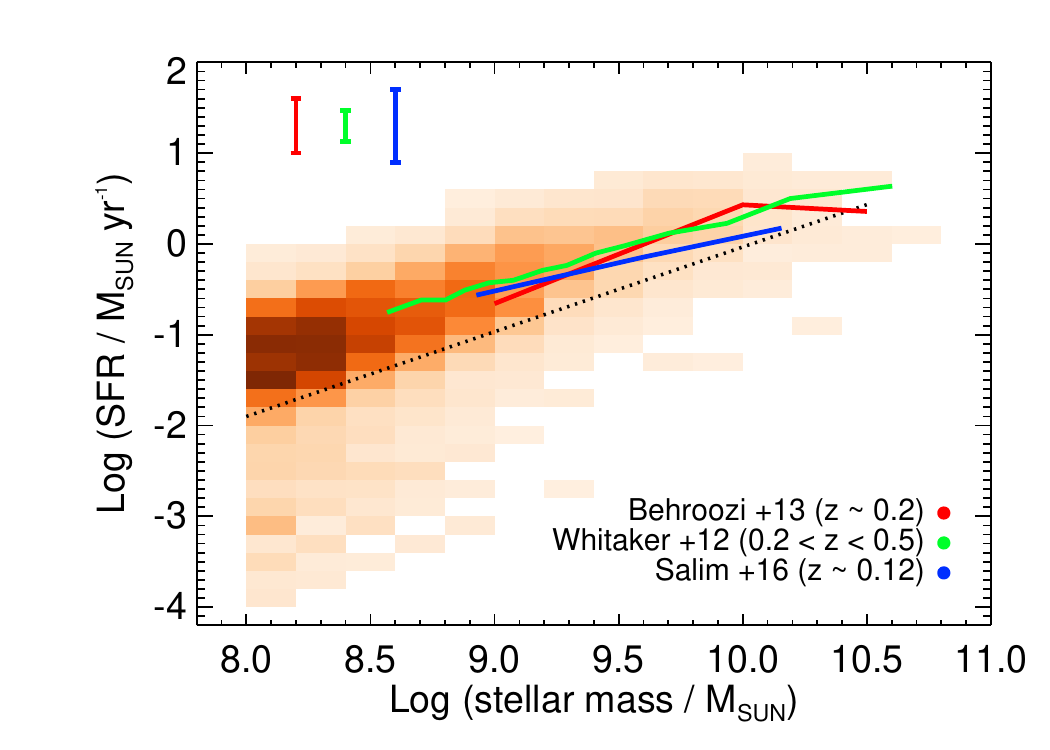}\\
\includegraphics[width=\columnwidth]{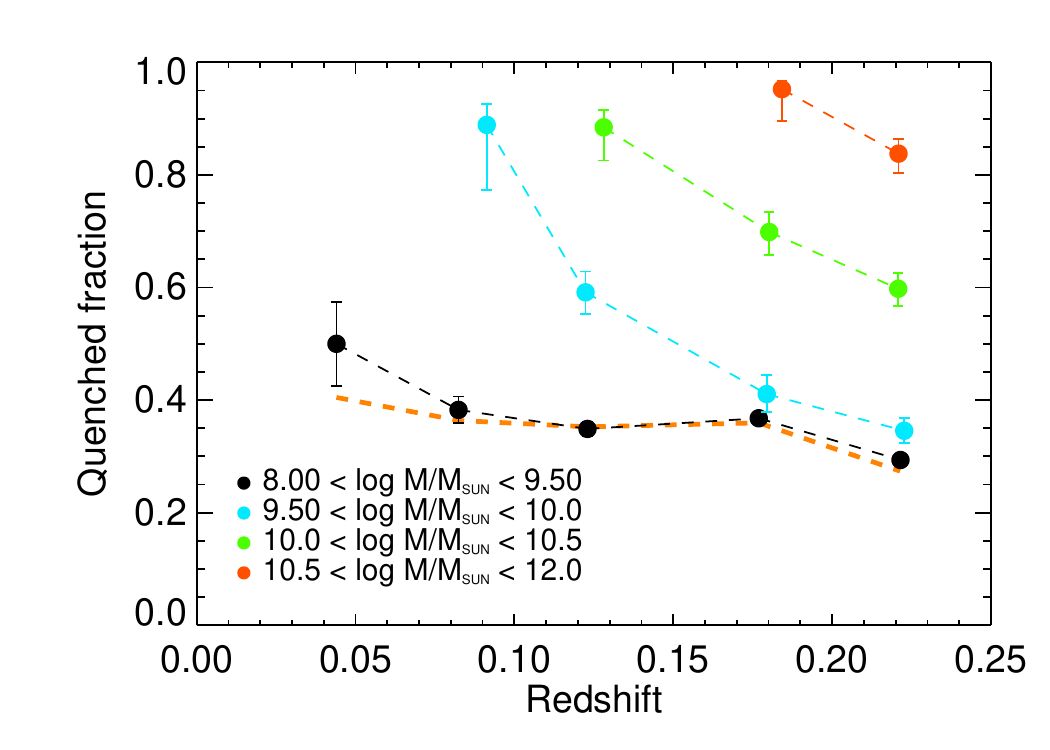}\\
\includegraphics[width=\columnwidth]{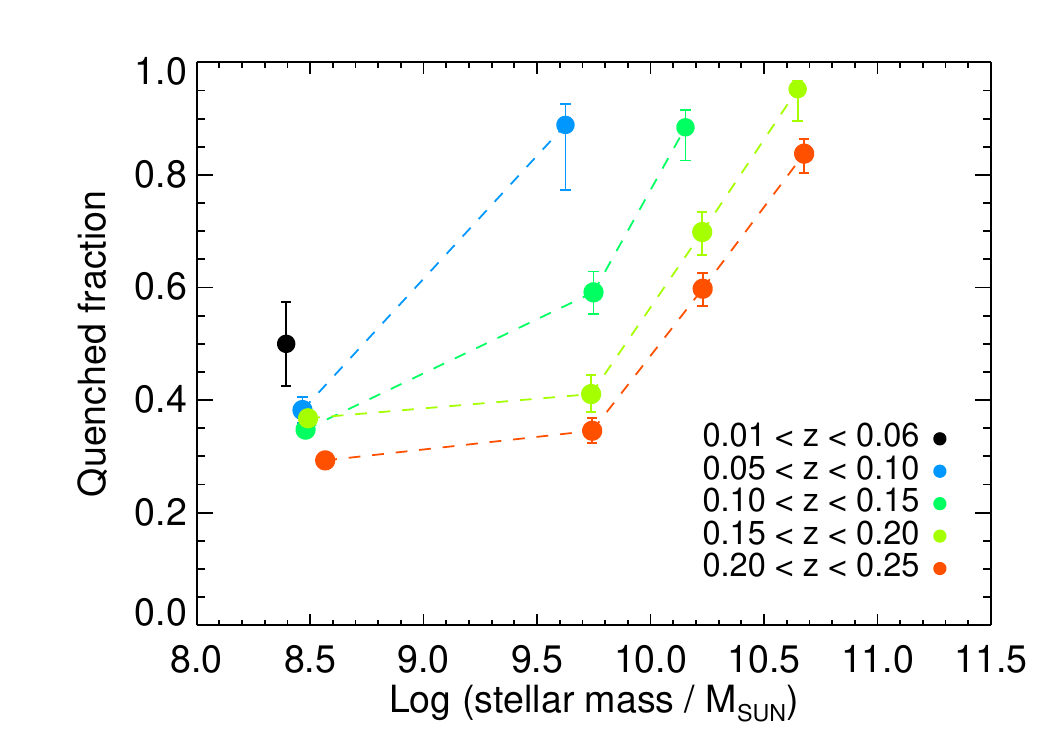}
\caption{\textbf{Top:} SFR vs stellar mass, in the nearby Universe ($z<0.25$) for the COSMOS2020 galaxy population (shown using the heatmap). The main well-defined locus, where SFR increases with stellar mass, is sometimes referred to as the star formation main sequence. The bottom of the SFMS, which is $\sim$0.5 dex below the main locus, is shown using the dotted line. SFMS measurements from the literature at low redshift, derived both using photometric and spectroscopic observations are shown overplotted and correspond well to their COSMOS2020 counterpart at relatively high stellar masses. The FWHMs of these main sequences are indicated in the upper left-hand corner. \textbf{Middle:} The quenched fraction (i.e. the fraction of galaxies that lie below the dotted line in the top panel), as a function of redshift, for different stellar mass ranges (see legend). The red fraction in dwarfs, from the right-hand panel in Figure \ref{fig:red_fraction_summary}, is shown overplotted using the dashed orange line. \textbf{Bottom:} The quenched fraction as a function of stellar mass for different redshift ranges (see legend).}
\label{fig:sfms_properties}
\end{figure}

\section{The star formation main sequence: quenched fractions as a function of redshift and stellar mass} 
\label{sec:sfms}

We proceed by exploring the star formation properties of our nearby dwarf galaxies. The top panel of Figure \ref{fig:sfms_properties} shows SFR vs stellar mass for our galaxy sample. The heatmap shows the COSMOS2020 galaxy population. The main well-defined ridgeline, where star formation rate increases with stellar mass, is sometimes referred to as the `star formation main sequence' (SFMS). The lower end of the SFMS, defined to be $\sim$ 0.5 dex below the ridgeline, is shown using the dotted line. To assess its veracity, we compare the COSMOS2020 ridgeline to other SFMS in the literature, derived using spectro-photometric data from the SDSS \citep[e.g.][]{Salim2016}, multi-wavelength photometric data based on medium-band photometry \citep{Whitaker2012} and the median of multiple SFMS measurements at $z\sim0.2$ \citep{Behroozi2013}. The full width at half maximum (FWHM) of the main sequences in each of these observational studies is indicated in the upper left-hand corner of the panel. The SFMS in our galaxies at relatively high stellar masses ($M_{\rm{\star}}$ > 10$^{8.5}$ M$_{\odot}$) agrees well with other SFMS. Given the improved detectability of dwarfs in the deep data used here, it appears that the SFMS continues uninterrupted with the same slope into the dwarf regime. 

The SFMS allows us to consider the recent star formation history in terms of the quenched fraction of galaxies, defined here as the fraction of galaxies which reside below the lower end of the SFMS (shown using the dotted black line in the top panel). In the middle and bottom panels of Figure \ref{fig:sfms_properties} we show the quenched fractions, both as a function of redshift for various stellar mass ranges (middle panel) and as a function of stellar mass for various redshift ranges (bottom panel). The red fraction in the dwarf regime, from Figure \ref{fig:red_fraction_summary}, is shown, for comparison, using the dashed orange line. Not surprisingly, the red and quenched fractions show a close correspondence, which indicates that the red fraction in the rest-frame $(g-i)$ colour could be used as a proxy for the quenched fraction. This is likely to be useful, for example, in cases where a large wavelength baseline may not be available, making it more difficult to estimate SFRs from SED fitting. $\sim$50 per cent of dwarfs are quenched at $z<0.05$, with the quenched fraction gradually falling to $\sim$30 percent by $z\sim0.25$.

\begin{figure*}
\center
\includegraphics[width=\columnwidth]{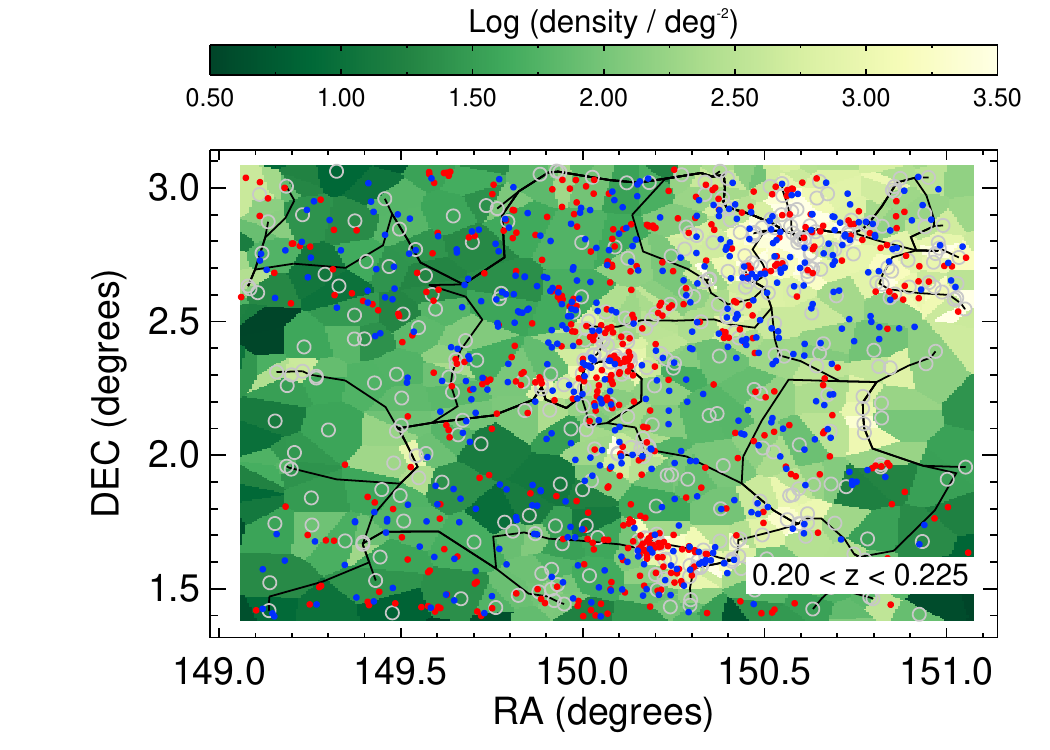}
\includegraphics[width=\columnwidth]{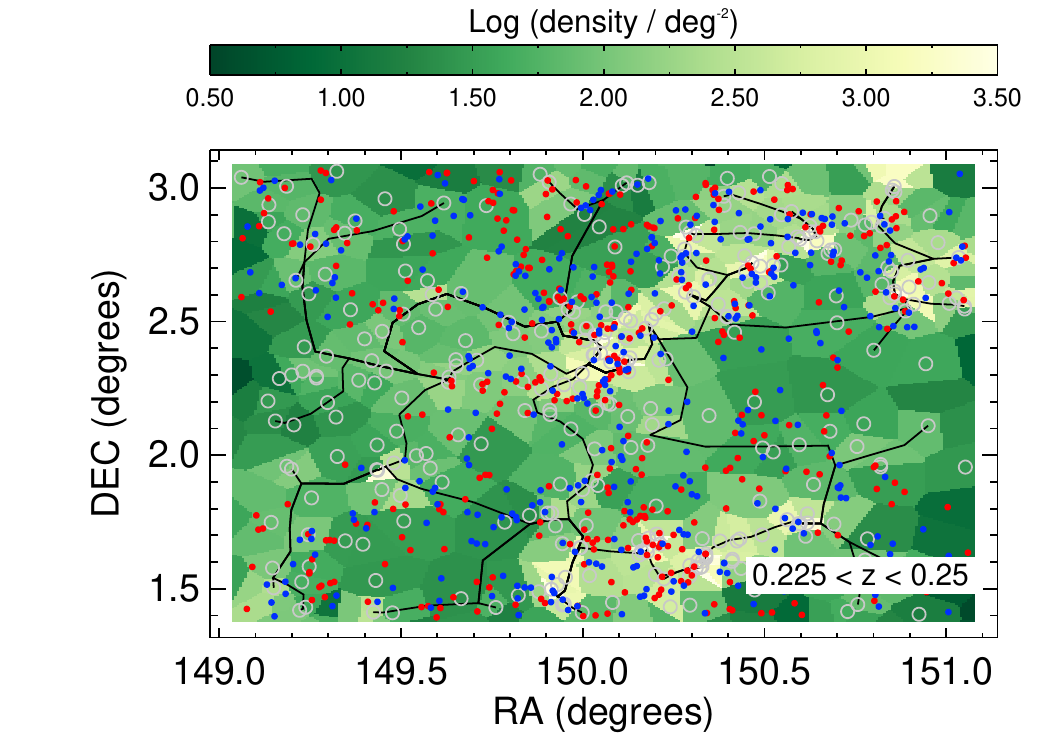}
\caption{Density maps in the COSMOS field at $0.2<z<0.25$ (created using DisPerSE), that are used for the density analysis in Section \ref{sec:quenching}. The colours indicate the local density (see legend), while the solid black lines show the locations of the filaments. The open grey circles show the positions of massive (M$_{\star}$ > 10$^{10}$ M$_{\odot}$) galaxies which are used for constructing the maps, as described in Section \ref{sec:disperse}. Red and blue dwarfs are shown using the filled red and blue circles respectively. 
The transverse size of the footprints are $\sim$25.7 (27.5) Mpc at $z=0.21 (0.23)$ respectively.}
\label{fig:density_map}
\end{figure*}

\begin{figure*}
\center
\includegraphics[width=\columnwidth]{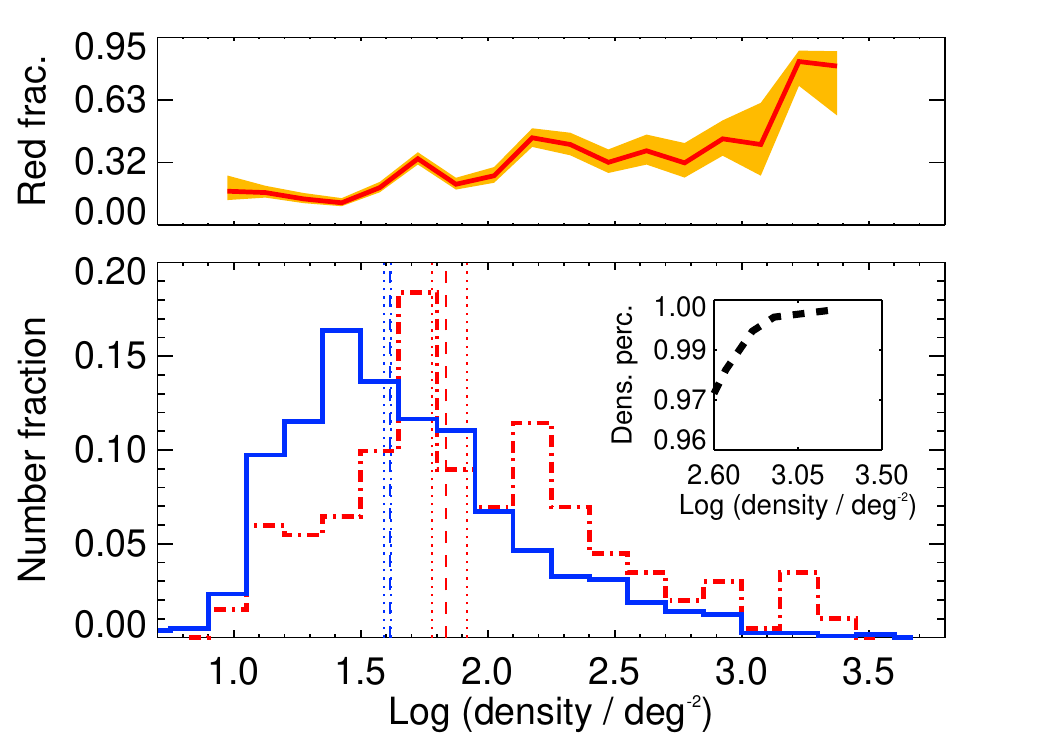}
\includegraphics[width=\columnwidth]{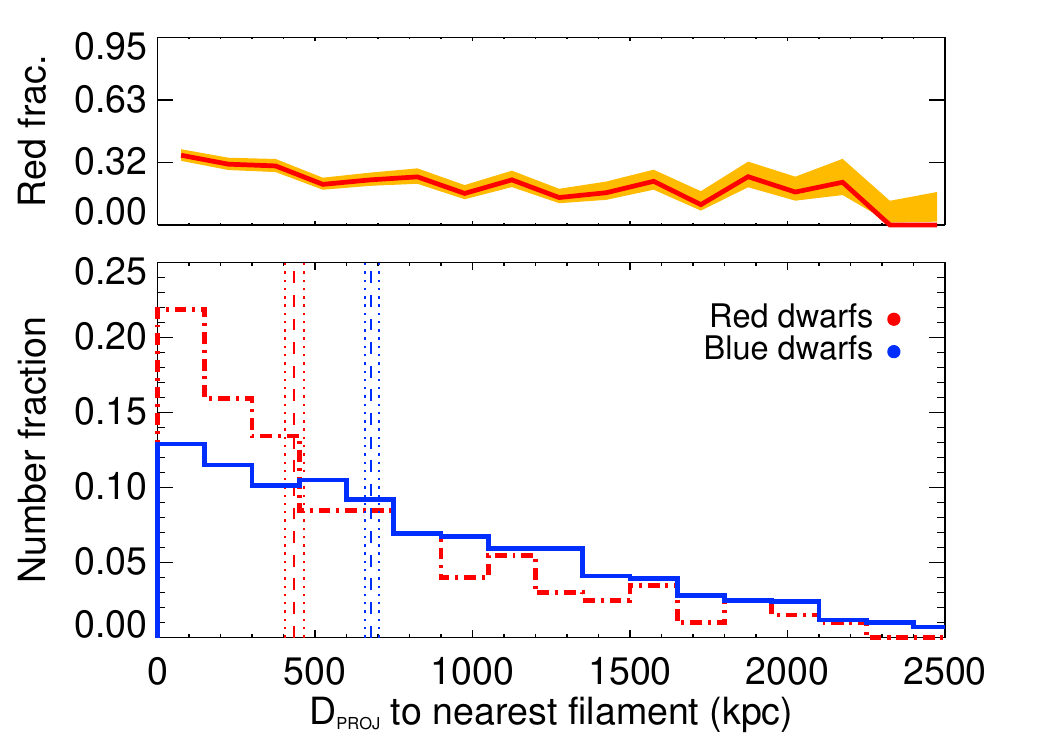}\\ \includegraphics[width=\columnwidth]{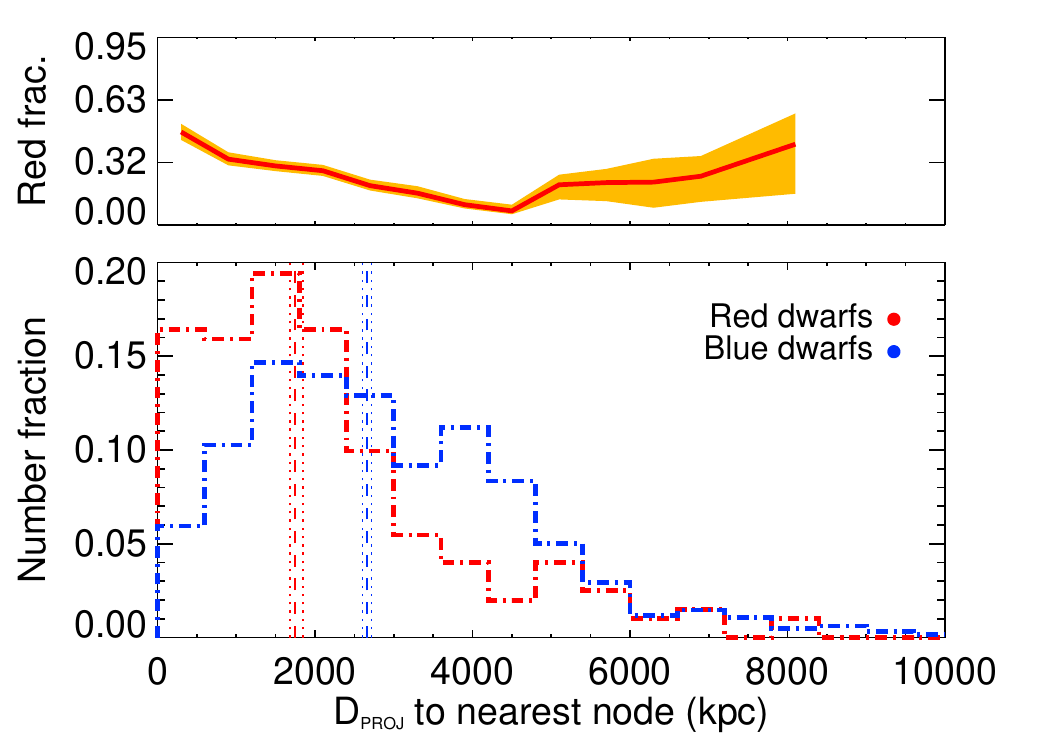} \includegraphics[width=\columnwidth]{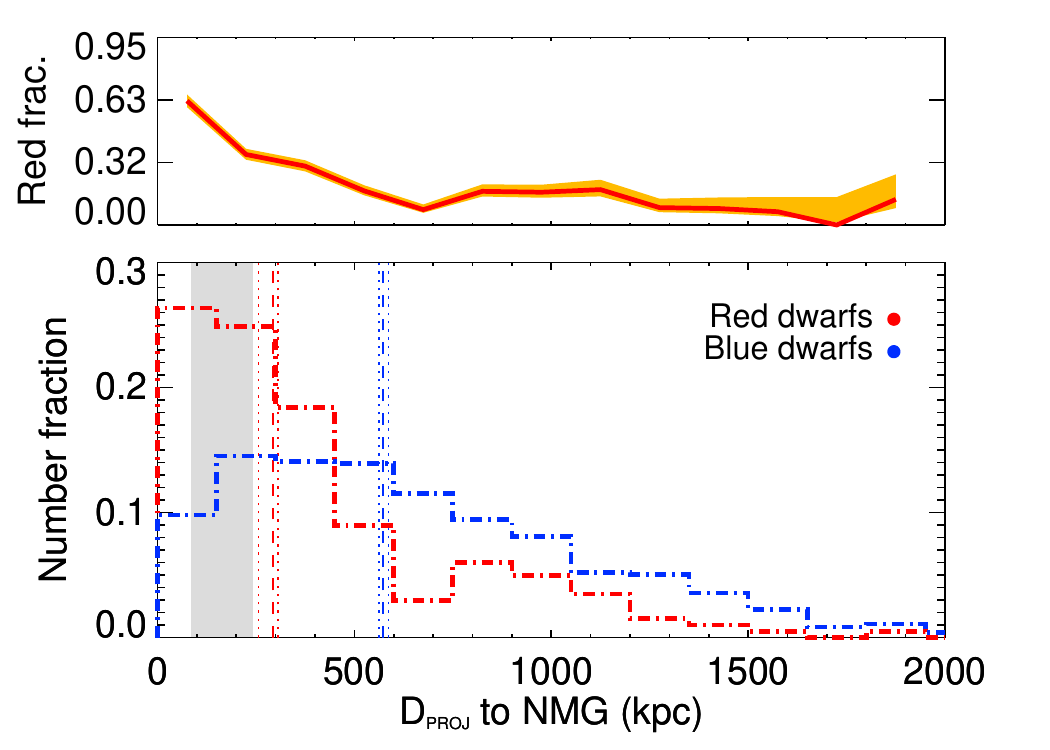}
\caption{\color{black}Properties of red and blue dwarfs as a function of environmental parameters. In each plot, the top panel shows the fraction of red dwarfs (red solid line) and the associated errors (orange regions), as a function of the environmental parameter of interest. The red fractions are only calculated when neither the number of red dwarfs nor the number of blue dwarfs is equal to zero. The bottom panel shows the distributions of red and blue dwarfs in the same environment parameter. We study the log density values (top-left) and the projected distances from the nearest filament (top right), the nearest node (bottom left) and the nearest massive galaxy (NMG; bottom right) for red and blue dwarfs. The inset in the top-left plot shows the density percentiles that correspond to the highest densities, where virtually all dwarfs become red. The shaded region in the bottom-right plot shows the range of values that corresponds to the virial radii of dark matter halos of massive ($M_{\star}$ > 10$^{10}$ M$_{\odot}$) galaxies from the Horizon-AGN simulation.}
\label{fig:distance_cosmic_web}
\end{figure*}


\section{What quenches dwarf galaxies?} 
\label{sec:quenching}

Our analysis above already offers some clues about the processes that quench dwarf galaxies. Figure \ref{fig:red_fraction_summary} indicates that the red fractions, in nearby dwarfs that reside in low-density environments, are $\sim$30 -- 40 per cent across our redshift range of interest. While the corresponding value in high-density environments like the core of Virgo is larger, the significant red fraction in low-density environments suggests that inhabiting a region where the ambient density is high is not a prerequisite for the quenching of star formation in dwarf galaxies. The probability of a dwarf being red may, therefore, correlate more closely with finer details of environment (e.g. proximity to a specific type of structure within the cosmic web) or be driven by internal processes, like baryonic feedback from stars and AGN. 

In this section, we probe the environments of our dwarfs in more detail, using the density maps described in Section \ref{sec:disperse}, to quantify the relative role of internal and external processes in quenching star formation in these systems. Recall from the discussion in Section \ref{sec:data} that, for this exercise, we consider the galaxy population in the redshift range $0.2<z<0.25$, where the redshift errors of massive ($M_{\star}$ > 10$^{10}$ M$_{\odot}$) galaxies, which are used to construct the density maps, reach a broad minimum, making our density maps most accurate. In addition, the number of massive galaxies is not large enough to construct density maps at $z<0.2$. 
Figure \ref{fig:density_map} shows two density maps covering the redshift range $0.2<z<0.25$, where regions of higher density are shown in lighter colours. The solid black lines show the skeleton i.e. the locations of the filaments. The open grey circles show the positions of the massive galaxies which are used to construct the density maps, while random samples of red and blue dwarfs are shown using the filled red and blue circles respectively. 

\begin{center}
\begin{table*}
\begin{tabular}{| l | c | c | c | c | c | c |}
\hline
\hline
& Red med. & Blue med. & Med. ratio (blue/red)  & Red FWHM & Blue FWHM & FWHM ratio (blue/red)\\
\hline
Log density (deg$^{-2}$) & 1.83 & 1.62 & 0.88 & 0.90 & 1.05 & 1.17 \\
Proj. dist. to nearest filament (kpc) & 420 & 682 & 1.63 & 415 & 876 & 2.11\\
Proj. dist. to nearest node (kpc) & 1611 & 2639 & 1.64 & 2515 & 4207 & 1.67\\
Proj. dist. to nearest massive galaxy (kpc) & 248 & 568 & 2.29 & 386 & 974 & 2.52\\
\end{tabular}
\caption{Comparison of red and blue dwarfs, in terms of their local density (row 1) and projected distances to the nearest filament (row 2), node (row 3) and massive galaxy (row 4). Columns 2-4 present median values for the red and blue dwarfs and the ratio of these medians (blue/red). Columns 5-7 present the same for the FWHMs of the distributions for the red and blue dwarfs. Red dwarfs typically inhabit regions of higher density and reside closer to nodes, filaments and massive galaxies. Their distributions also consistently show lower FWHMs i.e. they are more homogeneous in these properties as a population than the blue dwarfs.}
\label{tab:median_fwhm}
\end{table*}
\end{center}

{\color{black}In Figure \ref{fig:distance_cosmic_web}, we summarise the differences between red and blue dwarfs in terms of their local density (upper-left plot) and different aspects of the cosmic web. In the upper-right, lower-left and lower-right plots we consider the projected distances of red and blue dwarfs from the nearest filament, node and massive galaxy respectively. The top panel in each plot shows the fraction of red dwarfs, while the bottom panel shows the distributions of the red and blue dwarf populations. The inset in the upper left plot shows the density percentiles that correspond to the highest densities. The grey shaded region in the lower-right plot indicates the virial radii of dark matter halos that host massive (M$_*$ > $10^{10}$ M$_{\odot}$) galaxies in the local Universe, taken from the Horizon-AGN simulation\footnote{{\color{black}Horizon-AGN is a cosmological hydrodynamical simulation which offers a 100 $h^{-1}$ comoving Mpc$^3$ volume, with a spatial resolution (which is set by the gravitational softening) of 1 kpc in proper units, a dark matter mass resolution of $8\times 10^7$ M$_{\odot}$, a gas mass resolution of $\sim10^7$ M$_{\odot}$ and a stellar mass resolution of $4\times 10^6$ M$_{\odot}$ \citep{Dubois2014,Kaviraj2017}.}}}. 

In Table \ref{tab:median_fwhm} we present the median values and FWHMs of these distributions for the red and blue dwarfs. Figure \ref{fig:distance_cosmic_web} and Table \ref{tab:median_fwhm} confirm what is apparent from visual inspection of the density maps in Figure \ref{fig:density_map} i.e. that red dwarfs have a larger tendency of inhabiting regions of higher density and reside closer to nodes, filaments and massive galaxies. The distributions of the red dwarfs consistently show lower FWHMs i.e. they are more homogeneous in these properties as a population than blue dwarfs. $\sim$50 per cent of red dwarfs reside within projected distances that are less than the virial radii of massive galaxies (grey region in the lower-right plot within Figure \ref{fig:distance_cosmic_web}). There is no difference in either the average or the median density of the nearest filaments for red and blue dwarfs (not shown for brevity). The blue population virtually disappears at the very highest densities ($>3$ deg$^{-2}$), which correspond to the top 1 percentile in density values (see inset in the upper-left plot). This suggests that only at the highest density percentiles in the galaxy population that we study here does environmental quenching become strong enough to turn virtually all dwarfs red. 

The differences in the median properties between red and blue dwarfs (quantified using the ratio of the medians) is largest for the distance to the nearest massive galaxy. Proximity to a massive galaxy may therefore be the most important factor that influences whether a dwarf galaxy becomes red. The quenching of star formation due to the presence of a nearby massive galaxy is likely to be driven by the large tidal perturbations that dwarfs will encounter when they are within the virial radius of their massive neighbours \citep[e.g.][]{Jackson2021a}, which appears to be consistent with the large fraction ($\sim$50 per cent) of red dwarfs having projected distances less than the virial radii of massive galaxies. 

{\color{black}It is worth noting that the red fraction in dwarfs that reside at significant projected distances (e.g. $\sim$ 1 Mpc) from massive galaxies does not fall to zero (see the bottom right panel of Figure \ref{fig:distance_cosmic_web}). It is difficult to draw conclusions at even larger projected distances, given the large uncertainties that are driven by the number count errors. Noting that the 3D distances will be larger than their projected counterparts, our results appear inconsistent with a picture -- as suggested by recent work based on the SDSS -- in which dwarfs that reside at large distances from massive galaxies are never quenched.} 

Notwithstanding the broad differences in the locations of red dwarfs compared to their blue counterparts, it is worth noting that many red dwarfs actually exist in regions of low density. For example, around half of the red dwarf population resides at projected distances greater than the virial radii of massive galaxies. And around 15 (20) per cent of red dwarfs reside both outside the virial radii of massive galaxies and in regions which represent the lower 50 (70) per cent in density percentile. Comparison to the Horizon-AGN simulation suggests that dark matter halos at such density percentiles have masses less than 10$^{11.6}$ (10$^{12}$) M$_{\odot}$ respectively i.e. these regions have very low ambient galaxy density. 

A large fraction of red dwarfs must therefore be quenched by processes that are unrelated to environment. These are likely to be internal processes, such as stellar and AGN feedback \citep[e.g.][]{Volonteri2016,Kaviraj2019,Koudmani2021}, although the data at hand do not allow us to put detailed constraints on which internal mechanisms may dominate the quenching process. It does, however, highlight the need to improve our understanding of the impact, in dwarfs, of internal processes such as AGN feedback. While well-established as a potentially key driver of massive-galaxy evolution \citep[e.g.][]{Kaviraj2011,Beckmann2017,Kakkad2022}, the role of AGN is only beginning to be understood in the dwarf regime \citep[e.g.][]{Silk2017,Koudmani2022}. In summary, while local environment, particularly proximity to massive galaxies, appears to increase the probability of a dwarf being red, a large fraction of the quenching in the dwarf population as a whole is likely to take place via internal processes that are unrelated to local environment. 

{\color{black}We complete our study by comparing our findings to recent theoretical work. While quenched fractions in dwarfs from past shallow surveys may show some tension with those predicted by simulations, our findings are in good qualitative agreement with current theoretical models. It is worth noting first that precise comparisons are difficult because the definition of a quenched galaxy tends to vary between different studies. Nevertheless, the quenched fractions in dwarf galaxies in our stellar mass range of interest, which reside in low-density environments in the NewHorizon simulation \citep{Dubois2021}, are between 30 and 50 per cent (Rhee et al. in preparation). The results from the FIREbox simulation \citep{Feldmann2023} are similar, with around 50 per cent of dwarfs with stellar masses around 10$^8$ M$_{\odot}$ classified as quenched. Both simulations predict quenched fractions that are consistent with those that we have derived here. 

Interestingly, current simulations \citep[e.g.][]{Herzog2023,Pereirawilson2023} suggest that the quenched fraction in dwarfs that reside in low-density environments may be controlled largely by the ratio of halo and stellar mass, an internal characteristic which will influence the impact of processes like baryonic feedback, in qualitative agreement with our findings. In particular, quenched dwarfs in our stellar mass range of interest are predicted to be preferentially clustered around massive systems \citep[e.g.][]{Herzog2023}, which is strongly aligned with the results of our study.}


\section{Summary}
\label{sec:summary}

Dwarf galaxies dominate the galaxy number density, making them fundamental to a complete understanding of galaxy evolution. However, while they have been studied in our local neighbourhood (e.g. in the Local Group or around nearby massive galaxies), typical dwarfs are not bright enough to be detectable outside the very local Universe in past large surveys like the SDSS, which offer large footprints but are relatively shallow. The subset of dwarfs that is detected in such surveys have high SFRs which boost their luminosities, making them detectable in shallow images. However, this also makes them anomalously blue, a bias that likely affects the derivation of properties like red or quenched fractions and the identification of interesting sub-populations, such as galaxies that may host AGN.   

The advent of a new generation of surveys, that are both deep and wide, offers the possibility to construct, for the first time, unbiased statistical samples of dwarfs outside the local neighbourhood. These populations can be used to probe aspects of galaxy evolution in the dwarf regime that we were previously restricted to studying in massive galaxies only. Here, we have explored the properties of $\sim$7,000 dwarf (10$^8$ M$_{\odot}$ < $M_{\star}$ < 10$^{9.5}$ M$_{\odot}$) galaxies in the nearby Universe ($z<0.25$). The dwarf population in our study is complete down to $M_*$ $\sim$ 10$^{8}$ M$_{\odot}$, out to at least $z\sim0.3$. The galaxies that underpin this work are therefore unbiased across the stellar mass and redshift ranges explored here. At our redshifts of interest, the COSMOS field is not centred on high-density structures and is likely to be a low-density region containing galaxies in groups and the field. Our main conclusions can be summarised as follows: 

\begin{itemize}
    
    \item At low redshift ($z\sim0.05$), around 40 per cent of the dwarf population in low-density environments is red, with the red fraction decreasing with redshift to below 30 per cent by $z\sim0.25$. In comparison, the red fraction in the core of Virgo and around nearby massive galaxies (both of which are relatively high-density environments) are factors of 1.6 and 1.3 higher respectively. The corresponding red fractions calculated using the SDSS MGS are underestimated by around factors of 3 and 8 at $z<0.05$ and $z\sim0.08$ respectively. 


    \item The star formation main sequence appears to continue uninterrupted, in terms of its slope, into the dwarf regime, down to at least $M_{\rm{\star}}$ $\sim$ 10$^8$ M$_{\odot}$. 
            
    \item Around 50 per cent of dwarfs at $z\sim0.05$ are quenched i.e. lie below the star formation main sequence, with the quenched fraction falling gradually to around 30 percent by $z\sim0.25$. The red and quenched fractions track each other closely, suggesting that the red fraction may be a useful proxy for the quenched fraction in the dwarf regime at low redshift.
        
    \item Red dwarfs typically reside closer to nodes, filaments and massive galaxies. However, proximity to a massive galaxy appears to be more important in determining whether a dwarf is red, rather than simply its distance from the cosmic web or the mean density of its local environment.

    \item Notwithstanding these broad locational trends, many red dwarfs exist in regions of low density. For example, the red fraction does not fall to zero even when dwarfs are at significant ($\sim$ 1 Mpc) projected distances from massive galaxies. Around half of the red dwarfs reside at projected distances greater than the virial radii of massive galaxies. Furthermore, around 15 (20) per cent of the red dwarf population resides both outside the virial radii of massive galaxies and in regions which represent the lower 50 (70) per cent in density percentile (i.e. in regions of very low ambient density). A large fraction of red dwarfs are, therefore, quenched by mechanisms that are unrelated to environment, which are likely to be internal processes such as stellar and AGN feedback. 
    
\end{itemize}

In the coming years, new and forthcoming surveys like Euclid and LSST will provide unprecedented volumes of deep-wide multi-wavelength photometric data, of similar quality to what has been employed by this study. These datasets will enable the detailed exploration of stellar assembly, black-hole growth and morphological transformation in the dwarf galaxy regime, to the same precision as has hitherto been possible in massive galaxies. Such studies will drive important advances in our current understanding of galaxy evolution over a significant fraction of cosmic time. 


\section*{Acknowledgements}

{\color{black}We thank the anonymous referee for many constructive comments that enabled us to improve the original version of the paper.} SK, IL and AEW acknowledge support from the STFC (grant numbers ST/Y001257/1, ST/S00615X/1 and ST/X001318/1). SK also acknowledges a Senior Research Fellowship from Worcester College Oxford. IL acknowledges a PhD studentship funded by the Centre for Astrophysics Research at the University of Hertfordshire. CL acknowledges support from the Programme National Cosmology and Galaxies (PNCG) of CNRS/INSU with INP and IN2P3, co-funded by CEA and CNES.

The cutouts in Figure \ref{fig:different_surveys} were created using the cutout tool constructed by Yao-Yuan Mao which can be found at this address: https://yymao.github.io/decals-image-list-tool/. We are grateful to Yao-Yuan Mao for releasing this as public software.

The Hyper Suprime-Cam (HSC) collaboration includes the astronomical communities of Japan and Taiwan, and Princeton University. The HSC instrumentation and software were developed by the National Astronomical Observatory of Japan (NAOJ), the Kavli Institute for the Physics and Mathematics of the Universe (Kavli IPMU), the University of Tokyo, the High Energy Accelerator Research Organization (KEK), the Academia Sinica Institute for Astronomy and Astrophysics in Taiwan (ASIAA), and Princeton University. Funding was contributed by the FIRST program from the Japanese Cabinet Office, the Ministry of Education, Culture, Sports, Science and Technology (MEXT), the Japan Society for the Promotion of Science (JSPS), Japan Science and Technology Agency (JST), the Toray Science Foundation, NAOJ, Kavli IPMU, KEK, ASIAA, and Princeton University. This paper makes use of software developed for Vera C. Rubin Observatory. We thank the Rubin Observatory for making their code available as free software at http://pipelines.lsst.io/.

This paper is based on data collected at the Subaru Telescope and retrieved from the HSC data archive system, which is operated by the Subaru Telescope and Astronomy Data Center (ADC) at NAOJ. Data analysis was in part carried out with the cooperation of Center for Computational Astrophysics (CfCA), NAOJ. We are honored and grateful for the opportunity of observing the Universe from Maunakea, which has the cultural, historical and natural significance in Hawaii. This paper used data that is based on observations collected at the European Southern Observatory under
ESO programme ID 179.A-2005 and on data products produced by CALET and
the Cambridge Astronomy Survey Unit on behalf of the UltraVISTA consortium.

For the purpose of open access, the authors have applied a Creative Commons Attribution (CC BY) licence to any Author Accepted Manuscript version arising from this submission. 


\section*{Data Availability}
The data used in this study are taken from \citet{Weaver2022}. The density maps were created using the DisPerSE algorithm which is described in \citet{Sousbie2011}. 

\appendix

\section{Effective radii of dwarf galaxies in this study}
\label{sec:effective_radii}

{\color{black}Figure \ref{fig:hlr} shows the distribution of effective radii of the nearby dwarf galaxies in our study. The median seeing of the HSC images is $\sim$0.6 arcseconds. The median half-light radius is $\sim$0.78 arcseconds (show using the blue dashed line), while the value used to determine the completeness threshold in Section \ref{sec:deep_wide} is shown using the red dashed line. At the redshifts sampled here, the dwarf galaxies are relatively small which drives the similarities in the completeness curves derived using total magnitudes and effective surface brightnesses in Section \ref{sec:deep_wide}.} 

\begin{figure}
\center
\includegraphics[width=0.9\columnwidth]{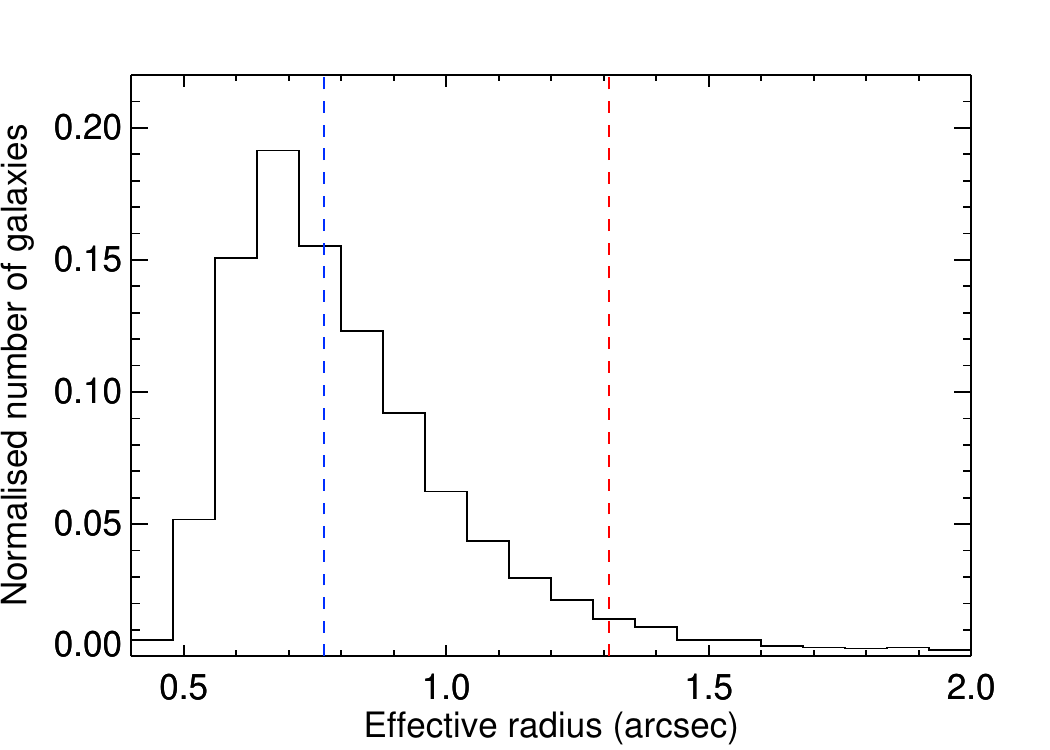}
\caption{Effective radii of our dwarf galaxies. The blue dashed line shows the median value, while the 95th percentile value, used to determine the completeness curves in Section \ref{sec:deep_wide}, is shown using a dashed red line. At our redshifts of interest, the dwarf galaxies studied here are small which drives the similarities in the completeness curves derived using total magnitudes and effective surface brightnesses in Section \ref{sec:deep_wide}.} 
\label{fig:hlr}
\end{figure}

\bibliographystyle{mnras}
\bibliography{references} 

\bsp	
\label{lastpage}
\end{document}